\documentclass[11pt,onecolumn,draftclsnofoot]{IEEEtran}

\usepackage[utf8]{inputenc}
\usepackage[english]{babel}
\usepackage{float}
\usepackage{amsmath}
\usepackage{amsfonts}
\usepackage{amssymb}
\usepackage{amsthm}
\usepackage{graphics}
\usepackage{comment}
\usepackage{algorithm,algorithmic}
\usepackage{array}
\usepackage{enumitem}
\usepackage[font={small, onehalfspacing}, labelfont=bf, skip=5pt, belowskip=0pt]{caption}
\usepackage{xcolor}
\usepackage{tikz}
\usepackage{tikz-network}
\usepackage{tcolorbox}
\usepackage{cuted}
\usepackage{lipsum}
\usepackage{multicol}
\usepackage{capt-of}
\usepackage{dblfloatfix}
\usepackage{multidef}
\usepackage{soul}
\usepackage{cite}
\usepackage{hyperref}
\hypersetup{
    colorlinks=true,
    linkcolor=blue,
    filecolor=magenta,      
    urlcolor=blue,
    pdftitle={Overleaf Example},
    pdfpagemode=FullScreen,
    }

\usetikzlibrary{colorbrewer}
\usetikzlibrary{arrows}

\definecolor{tab1}{HTML}{1f77b4}
\definecolor{tab2}{HTML}{ff7f0e}
\definecolor{tab3}{HTML}{2ca02c}
\definecolor{tab4}{HTML}{d62728}
\definecolor{tab5}{HTML}{9467bd}
\definecolor{tab6}{HTML}{8c564b}
\definecolor{tab7}{HTML}{e377c2}

\newcommand{\red}[1]{\textcolor{red}{#1}}

\multidef[prefix=set]{\ensuremath{\mathbb{#1}}}{A-Z}

\multidef[prefix=cal]{\ensuremath{\mathcal{#1}}}{A-Z}

\multidef[prefix=v]{\ensuremath{\mathbf{#1}}}{a-z}

\multidef[prefix=m]{\ensuremath{\mathbf{#1}}}{A-Z}

\begin{document}

\title{From Nano to Macro: Overview of the IEEE Bio Image and Signal Processing\\ Technical Committee}

\author{Selin Aviyente,~\IEEEmembership{Senior Member, IEEE}, Alejandro Frangi,~\IEEEmembership{Fellow, IEEE}, Erik Meijering, ~\IEEEmembership{Fellow, IEEE}, Arrate Muñoz-Barrutia,~\IEEEmembership{Senior Member, IEEE}, Michael Liebling,~\IEEEmembership{Member, IEEE}, Dimitri Van De Ville,~\IEEEmembership{Fellow, IEEE}, Jean-Christophe Olivo-Marin,~\IEEEmembership{Fellow, IEEE}, Jelena Kovačević,~\IEEEmembership{Fellow, IEEE}, Michael Unser,~\IEEEmembership{Fellow, IEEE}
\thanks{The authors are the present and past BISP TC Chairs in reverse chronological order.}
\thanks{S. Aviyente is with the Department of Electrical and Computer Engineering, Michigan State University, East Lansing, MI, 48824 USA, Email: aviyente@egr.msu.edu.}
\thanks{Alejandro F Frangi is with the Centre for Computational Imaging and Simulation Technologies in Biomedicine (CISTB), Schools of Computing and Medicine, University of Leeds,  LS2 9JT Leeds, UK; the Alan Turing Institute, London, UK; and the Departments of Electrical Engineering (ESAT) and Cardiovascular Sciences, KU Leuven, Leuven, Belgium, Email: a.frangi@leeds.ac.uk.}
\thanks{Erik Meijering is with the School of Computer Science and Engineering, University of New South Wales, Sydney, NSW 2052, Australia. Email: erik.meijering@unsw.edu.au.}
\thanks{Arrate Muñoz-Barrutia is with the Bioengineering Department, Universidad Carlos III de Madrid, Leganés, Madrid, Spain. Email: mamunozb@ing.uc3m.es}
\thanks{Michael Liebling is with the Idiap Research Institute, Martigny, Switzerland. Email: michael.liebling@idiap.ch}
\thanks{Dimitri Van De Ville is with the Neuro-X Institute, School of Engineering, EPFL, Geneva and Faculty of Medicine, University of Geneva. Email: dimitri.vandeville@epfl.ch}
\thanks{Jean-Christophe Olivo-Marin is with the Cell Biology and Infection Department, Institut Pasteur, Paris, France. Email: jcolivo@pasteur.fr}
\thanks{Jelena Kovačević is with the New York University Tandon School of Engineering, Brooklyn, NY. Email: jelenak@nyu.edu.}
\thanks{Michael Unser is with the School of Engineering, EPFL, Lausanne, Switzerland. Email: Michael.Unser@epfl.ch.}
}

\maketitle

\section{Introduction}
The Bio Image and Signal Processing (BISP) Technical Committee (TC) of the IEEE Signal Processing Society (SPS) promotes activities within the broad technical field of biomedical image and signal processing. Areas of interest include medical and biological imaging, digital pathology, molecular imaging, microscopy, and associated computational imaging, image analysis, and image-guided treatment, alongside physiological signal processing, computational biology, and bioinformatics. BISP has 40 members and covers a wide range of EDICS, including CIS-MI: Medical Imaging, BIO-MIA: Medical Image Analysis, BIO-BI: Biological Imaging, BIO: Biomedical Signal Processing, BIO-BCI: Brain/Human-Computer Interfaces, and BIO-INFR: Bioinformatics. BISP plays a central role in the organization of the IEEE International Symposium on Biomedical Imaging (ISBI) and contributes to the technical sessions at the IEEE International Conference on Acoustics, Speech and Signal Processing (ICASSP), and the IEEE International Conference on Image Processing (ICIP). In this paper, we provide a brief history of the TC, review the technological and methodological contributions its community delivered, and highlight promising new directions we anticipate.

\section{Historical context}
\begin{figure}[t]
\centering
% \begin{subfigure}[b]{0.45\textwidth}
    \includegraphics[width=0.5\textwidth]{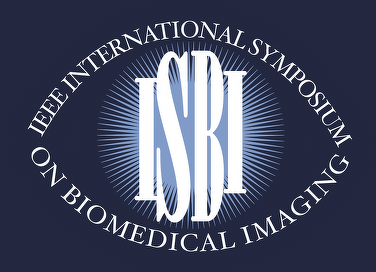}
% \end{subfigure}
% \hfill
% \begin{subfigure}[b]{0.45\textwidth}
%     \includegraphics[width=\textwidth]{ISBI_stats.png}
%     \caption{Paper statistics of ISBI 2012-2022}
%     \label{fig:stats}
% \end{subfigure}
\caption{Original ISBI logo designed by Annette Unser (graphic artist and sister of the founding chair), with the eye projecting a distinctive vision for ISBI. Observers have suggested that the central motif illustrates the Fourier slice theorem; or for the more pessimistic ones, the typical artifacts of the filtered back-projection reconstruction algorithm.}
\label{fig:ISBI}
\end{figure}

Until 2002, the signal processing activities related to biomedical imaging were overseen by the Image and Multidimensional Digital Signal Processing (IMDSP) committee of IEEE SPS, and typically presented in topical sessions at ICIP and ICASSP. IEEE SPS also co-sponsored the IEEE Transactions on Medical Imaging. Yet, at the turn of the century, the importance of imaging in medicine and biology was becoming increasingly apparent. At the same time, advanced signal processing played an ever-increasing role in the reconstruction and analysis of the vast volume of images produced. This realization was reinforced by the creation of the National Institute of BioImaging and Bioengineering (NIBIB) by the NIH and US Congress in December 2000, as an agency solely dedicated to the advancement of imaging technology and bioengineering. The latter was an official recognition of the crucial role of engineering in biomedical research and of the necessity to fund such research activities. This motivated the IEEE SPS and the IEEE Engineering in Medicine and Biology Society (EMBS) to join forces and demonstrate leadership in biomedical imaging research. Accordingly, it was decided to launch a new regular meeting on biomedical imaging: the IEEE International Symposium on Biomedical Imaging (ISBI) (Figure~\ref{fig:ISBI}), in close collaboration with NIBIB. The task of organizing this conference was given to Michael Unser (SPS representative) and Zhi-Pei Liang (EMBS representative). The fact that Prof. Unser had spent the larger part of his career at the NIH facilitated the interaction with NIBIB, which committed to supporting the first edition of ISBI
% redundant (with a grant of \$30k to organize special sessions and sponsor student participation), 
that took place in Washington, DC, USA, in July 2002. The unique aspect of ISBI was to cover the whole spectrum and range of imaging, from nano (electron and optical microscopy) to macro (medical imaging modalities) \cite{unser2002guest}.

\subsection{Creation of a Dedicated Technical Committee}
With the creation of ISBI and its establishment as the IEEE flagship conference in biomedical imaging, the next step was to put in place a structure to promote the conference, and ensure its scientific quality. Since Prof. Unser with his team had formulated the vision for ISBI, he was instructed to form an SPS TC on Bio Image and Signal Processing and to make suggestions for its initial membership. In addition to its strategic role in bioimaging, BISP was given the mission to oversee the SPS activities in biomedical signal processing (e.g., analysis of physiological signals) and bioinformatics---in short, to be responsible for all signal processing activities in medicine and biology, and to maintain a liaison with its sister TC in EMBS on Biomedical Imaging and Image Processing (BIIP). Since the inception of the TC, BISP members have also actively participated in cross-society activities, such as the IEEE Life Science Technical Community (LSTC) and the IEEE BRAIN Technical Community.

\subsection{Workshops and Conferences}
The inaugural ISBI was held between July 7-10, 2002, at the Ritz-Carlton Hotel, Washington, DC. The meeting was organized jointly by the SPS and the EMBS. Significant support was provided by both NIH and NIBIB ($\$40$K in grants and approximately 50 paid registrants). The conference was a huge success, providing a venue for interdisciplinary exchange with researchers from both medical and biological imaging areas. It was also well attended by NIH representatives. Dr. Elias Zerhouni gave the opening address, the then newly appointed NIH director. This attracted many observers, including members of the press. Dr. Roderic Pettigrew also delivered a speech---the very first in his new function as director of NIBIB. Both directors expressed a strong interest in the conference and commented on the need to strengthen the links between the engineering and biomedical research communities.

ISBI 2002 had two parts to the scientific program: 1) the contributed papers reviewed by the technical program committee, and 2) the invited papers. Out of 355 submitted papers, 73 were accepted for oral presentation and 142 for poster presentation. The invited program consisted of 10 special sessions that were organized by leading researchers in the field.

\begin{figure}
    \centering
    \includegraphics[width=\textwidth]{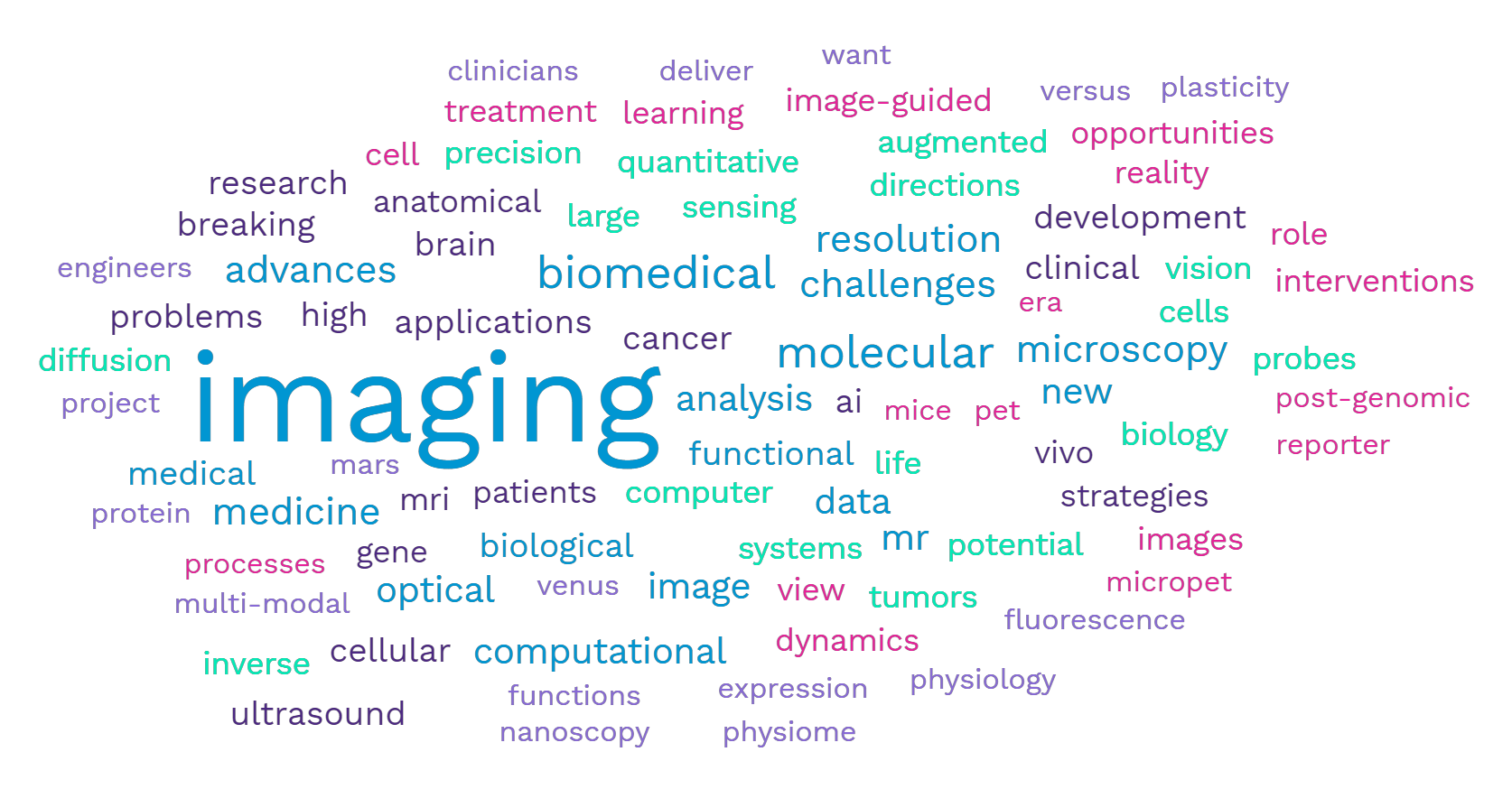}
    \caption{Word cloud with the most frequent keywords from the titles of all the ISBI keynotes in the past 20 years. Clinicians, Nobel Prize Winners, and NIH Officials have given these talks, among other contributors. }
    \label{fig:isbi-keynotes}
\end{figure}

Despite a very active and growing community, BISP membership has only increased moderately, yet the TC has always strived for a well-balanced representation across the broad range of sub-fields it covers. A key task for BISP members was to ensure that papers submitted to ISBI (or to dedicated tracks at ICASSP and ICIP) would benefit from the availability of a highly qualified pool of reviewers and editors. BISP members also participated in the many activities related to ISBI, as members of the organizing committee. Since 2006, ISBI has been held regularly as an annual four-day conference. Figure \ref{fig:isbi-keynotes} summarizes with a word cloud the keywords from the keynote titles since ISBI's inception. Outstanding clinical and technical speakers delivered their visions for the field, relevant trends, or challenges ahead. Among our distinguished speakers, there were Nobel Prize winners and top NIH officers. In 2022, ISBI was held for the first time as a fully hybrid conference in Kolkata, India, with every session having both physical and online speakers and audiences. Out of 785 submitted papers, 309 were accepted. In addition to the regular paper sessions, there were 5 special sessions, 5 plenary talks, 6 challenges, and 6 tutorials. In addition to ISBI, BISP has been an active contributor to ICASSP since 2006, with the number of submitted papers increasing from 100 in 2006 to 222 in 2020.

\section{Biomedical Image and Signal Acquisition Across Scales}
\begin{figure*}[t]
\centering
\includegraphics[width=\textwidth]{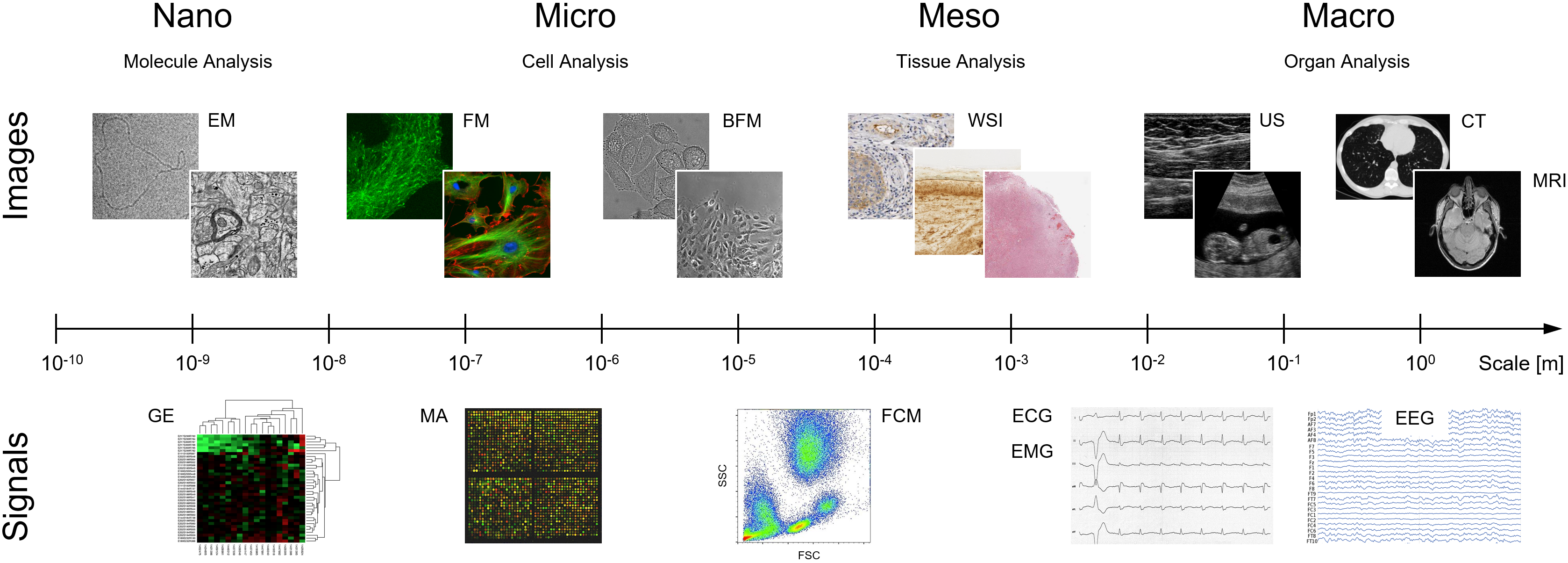}
\caption{Examples of the many data acquisition modalities in biomedical image and signal processing operating at various spatial scales. Abbreviations: BFM = bright-field microscopy, CT = computed tomography, ECG = electrocardiography, EEG = electroencephalography, EMG = electromyography, FCM = flow cytometry, FM = fluorescence microscopy, GE = gene expression, MA = micro array, MRI = magnetic resonance imaging, US = ultrasound, WSI = whole-slide imaging.}
\label{fig:modalities}
\end{figure*}

Biomedical data comes in many shapes and forms. BISP focuses on digital images and signals, which can be automatically processed and analyzed by advanced computational methods. To study biological processes in health and disease, many images and signal acquisition techniques have been developed in the past century, reflecting the fact that biological phenomena occur at different spatial and temporal scales (Figure~\ref{fig:modalities}). Before discussing methodological advances, we survey some of the most prominent modalities for molecular and cellular imaging, tissue imaging, anatomical and functional medical imaging, neuroimaging, physiological signal recording, and several data types in bioinformatics.

\subsection{Molecular and Cellular Imaging}
Molecular and cellular imaging has undergone multiple revolutions in the past three decades, moving from a mainly qualitative to a mostly quantitative field thanks to advances in molecular probes as well as imaging modalities \cite{murphy2005, kovacevic2006, munoz2015quantitative}. With the advent of the green fluorescent protein (GFP), pioneered by Osamu Shimomura, Martin Chalfie, and Roger Y. Tsien (Nobel Prize in Chemistry 2008), microscopy has become one of the key tools in biological research \cite{kovacevic2006}. Fluorescence microscopy became a fast-growing field to study, quantitatively and often within a high throughput/content setup, processes, and organelles within living cells and organisms. More recently, a vast leap has been made with the invention of super-resolution techniques, based on seminal work by Eric Betzig, Stefan W.\ Hell, and William E.\ Moerner (Nobel Prize in Chemistry 2014). Another recent development is selective plane illumination (light sheet) microscopy, which allows long-term biological studies of living organisms with a rapid acquisition, high resolution, and minimal phototoxicity. Classical image and signal processing methods, as well as modern deep learning-based methods, are increasingly used not only for reconstruction and deconvolution of the data produced by advanced microscopy imaging modalities, but also for enabling downstream tasks such as segmentation, classification, tracing, and tracking \cite{munoz2015quantitative, kervrann2016, meijering2022deep}. Fluorescence microscopy has enabled the study of dynamic processes within cells and complements structural and static imaging modalities, such as scanning probe microscopy, electron microscopy (Nobel Prize in Physics 1986), cryo-electron microscopy (Nobel Prize in Chemistry 2017), which have become part of the vast arsenal of tools for the life sciences. In recent years, several new journals or sections in established publications, e.g., Biological Imaging, Frontiers in Bioinformatics, Cell Reports Methods, have been launched to host the increasing number of publications in this domain.

\subsection{Tissue Imaging}
Microscopy can also be employed to study biological phenomena at the tissue level. Rather than imaging individual cells (molecular/cellular imaging) for basic research, or anatomical structures and entire organs (medical imaging) for clinical diagnostics, imaging of tissue slides prepared from biopsies enables characterizing and grading disease processes ex vivo as revealed by abnormal cell arrangements and tissue architectures. This is especially important in researching and diagnosing pathologies, notably the many types of cancer (the field of oncology), known to manifest themselves first at the cell and tissue level (histopathology). Recent advances in digital whole-slide imaging (WSI) systems (sometimes referred to as virtual microscopy) have created unprecedented opportunities for computer-aided diagnosis (CAD) in histopathology. Image and signal processing play a prominent role in histopathological image analysis, especially for breast cancer, prostate cancer, lung cancer, tumor pathology in many other forms of cancer, and cancer prognosis. Review papers have summarized and commented on the challenges and opportunities in this domain \cite{Gurcan-2009, Zhou-2020}. Typical tasks include the detection and segmentation of cell nuclei, glands, and lymphocytes, and computing various quantitative morphological features for classification. This, in turn, requires effective techniques for image normalization as well as feature selection and dimensionality reduction. Analysis of the spatial arrangements of tissues is often facilitated by graph-based representation and topological modeling. The challenges in histopathological image analysis are not only due to the high complexities of the image structures, but also to the typically large image sizes, on the order of tens of thousands by tens of thousands of pixels, at multiple magnifications. Traditionally, tissue classification has been performed using handcrafted features and machine learning methods, such as support vector machines (SVM) and random forests (RF). Still, there is now growing evidence that deep artificial neural networks provide fast and reliable image analysis on par with seasoned pathologists and can serve as a synergistic tool for the latter to improve accuracy and throughput. However, the full adoption of deep learning methods in pathology is hindered by the lack of large and reliably annotated image cohorts documenting the large diversity of diseases and the high variability of disease traits, calling for efficient automated annotation methods.

\subsection{Medical Imaging} Medical imaging refers to the imaging techniques and processes to gain insights into the interior of a body for clinical diagnosis or medical intervention, as well as visual representation of the function of organs or tissues. Medical imaging can be divided into structural or anatomical imaging and functional or physiological imaging. Many medical imaging techniques have been invented since the discovery of X-rays by Wilhelm Conrad R{\"o}ntgen in 1895, which form the basis of projection X-rays and computed tomography. In 1946, Bloch and Purcell (Nobel Prize 1952) independently discovered nuclear magnetic resonance, which formed the basis of MRI (Paul Lauterbur and Sir Peter Mansfield in the 70s, Nobel Prize 2003). MRI developed into a platform technology with many specialized techniques, providing insights into anatomy, perfusion, diffusion, or deformation. Ultrasound was used in medicine since World War II, but it was not until the late 70s that ultrasound imaging was popularized as a clinical imaging modality. In 1963, David Kuhl and Roy Edwards introduced emission reconstruction tomography, a method that later became single-photon emission computed tomography (SPECT). Sir Godfrey Hounsfield (Nobel Prize 1979) developed the first prototype of a computed tomography (CT) scanner in 1963, thanks to the availability of modern computers to solve the complex image reconstruction problems that ensued. Michael Hoffman, Michel Ter-Pogossian and Michael E. Phelps built the first positron emission tomography (PET) camera in 1974. Underpinning these techniques, there are considerable signal and image processing problems, ranging from image reconstruction, image deconvolution, image denoising and restoration, image transformation, and multimodal image co-registration. Over the last 3-4 decades, several signal processing developments had major impacts on medical imaging. For example, wavelets and splines played a major role in medical image interpolation, denoising, and filtering. Mutual information and other information theoretic metrics revolutionized multimodal image registration. Compressed sensing provided a novel approach to find solutions to under-determined linear equations with major impact in image reconstruction from projections (CT), from k-space (MRI) or from sensors (US), and, of course, the latest developments of deep learning and their impact across the board.

\subsection{Neuroimaging: From Images to Connectomes} 
\begin{figure}[t]
\centering
\includegraphics[width=\textwidth]{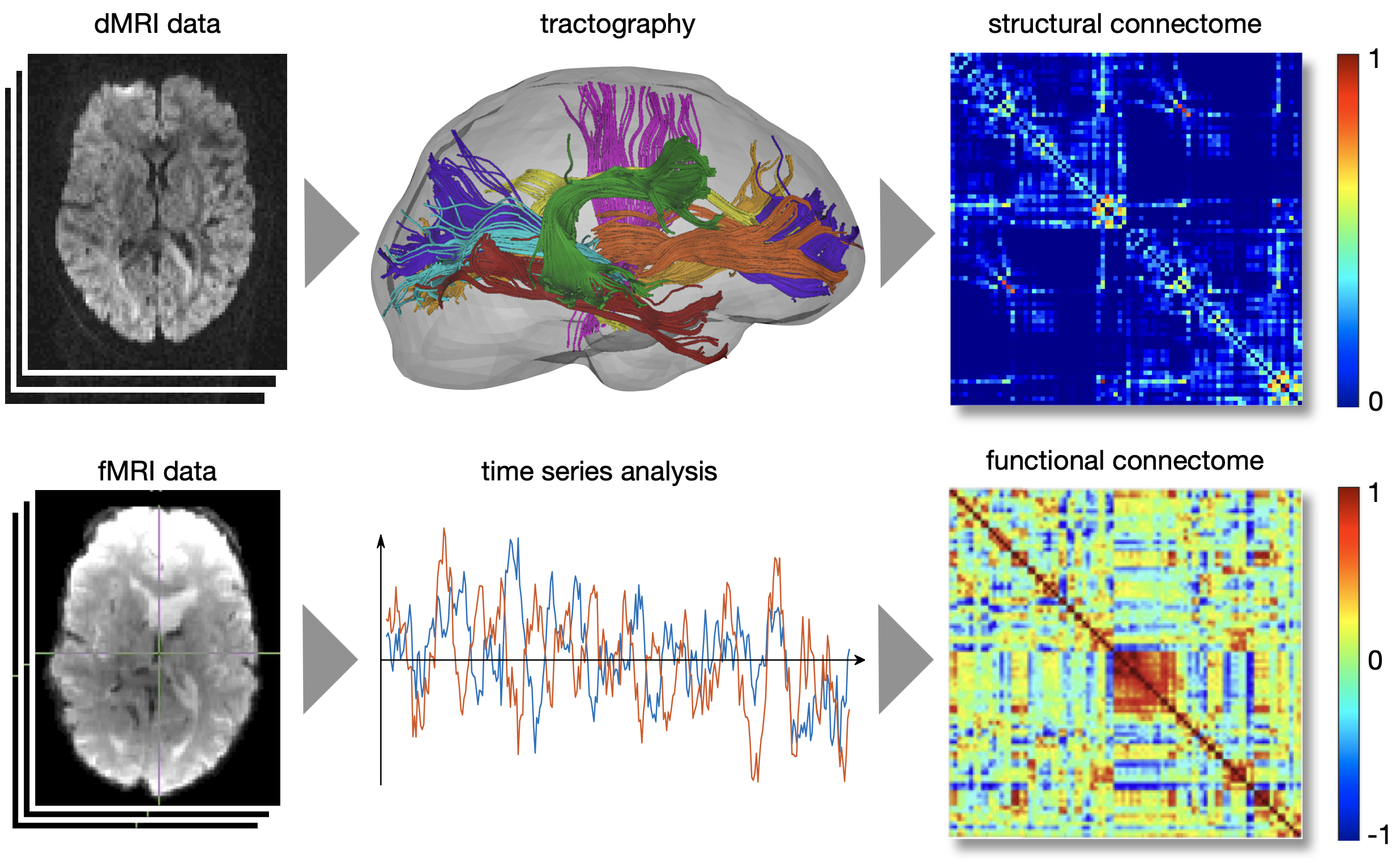}
\caption{Structural and functional connectomes play a key role in representing relationships between brain regions. Top: From diffusion-weighted MRI (dMRI), the orientation of axonal bundles in white matter can be extracted and processed by tractography to obtain the strength of structural connectivity between all pairs of regions. Bottom: Functional MRI (fMRI) provides a series of volumes where the blood oxygenation-level dependent (BOLD) signal is related to neuronal activity. Time series analysis exists in large diversity, but the functional connectome that reflects the statistical inter-dependencies between pairs of time series is one of them.}
\label{fig:connectomes}
\end{figure}

Another flourishing outlet for biomedical signal and image processing has been the processing of structural and functional MRI (fMRI) data~\cite{adali2008introduction}. The concept of establishing connectivity between brain regions has been fundamental in many emerging methodologies~\cite{van2016introduction} (Figure~\ref{fig:connectomes}). Structural connectivity is defined by the strength of inter-regional axonal fiber pathways that can be revealed using tractography methods applied to diffusion-weighted MRI (dMRI) data. Functional connectivity relates to the statistical interdependency between two time series of blood oxygenation-level dependent (BOLD) activity. When established for all possible pairs of regions, defined by a brain atlas, this leads to the structural and functional connectomes, respectively. While hemodynamic imaging has been the most commonly used modality for constructing the functional connectome, neurophysiological signals such as magneto/electroencephalography (M/EEG) have also been adopted thanks to their high temporal resolution. Both model-based and data-driven methods have been developed to quantify functional connectivity, including multivariate auto-regressive models, graphical models, phase synchrony and information-theoretic metrics.
Functional connectivity is also intimately related to blind source separation. This, in turn, relies on decomposing the data matrix into components driven by maximizing covariance (using techniques such as principal component analysis (PCA), singular value decomposition (SVD), or higher-order SVD (HOSVD)) or statistical independence (using independent component analysis (ICA)), which has become part of the pipeline in fMRI software suites. During the past decade, dynamic functional connectivity has been introduced to acknowledge the changing patterns of co-fluctuations, either by sliding-window functional connectivity, instantaneous activation patterns, or auto-regressive models. 

The resulting connectomes are commonly represented by graphs and analyzed to reveal organizational principles using, for instance, local clustering coefficient, efficiency, small-worldness, centrality, and phenotype behavior and disorder. These approaches have been applied to other species’ connectomes obtained using different modalities, leading to a new field, network neuroscience, which further branched out to machine learning, information theory, and computational neuroscience. Finally, the emergence of graph signal processing has found its way into the neuroimaging field~\cite{Huang2018}, providing a way to combine brain structure (i.e., graph defined by the structural connectome) and brain function (i.e., graph signals obtained by fMRI snapshots of brain activity).

\subsection{Physiological Signal Processing (M/EEG, EMG, ECG)} With the advent of wearable sensors, physiological signals are collected for various applications and are central to multiple new technologies, including brain computer interfacing (BCI) / human-machine interfacing (HMI), neurorehabilitation, neuroprosthetics, in addition to medical diagnosis and monitoring \cite{farina2021signal}. The most employed physiological signals include electromyogram (EMG), respiration, speech, heart rate variability, photoplethysmography (PPG), electrocardiogram (ECG) and magneto/electroencephalogram (M/EEG). Some prominent application areas that routinely rely on physiological signals include emotion recognition, autonomous driving, mental health and assistive technologies. For example, physiological changes such as heart rate, skin conductance and PPG signals are monitored for measuring human emotions as they are more reliable and harder to alter compared to explicit behaviors such
as facial expressions and speech. Similarly, the design of human-machine interfacing systems requires the consistent and accurate
decoding of motor intent with minimal
training and calibration. The multimodal, high-density sensing technology coupled with the nonstationary and nonlinear nature of biological signals requires the development of innovative signal processing and machine learning techniques to process, decompose, and decode these signals. Some methodologies employed in this area of research include blind source separation, time-frequency analysis, multimodal data fusion, supervised (or semi-supervised) learning and deep learning.  Different tasks, such as event detection, prediction and diagnosis, have been addressed using these tools.  

\subsection{Bioinformatics}
Since the turn of the century, major advances in molecular biology, along with advances in genomic data acquisition technologies, led to to the growth of biological data generated and shared by the scientific community; e.g., The Cancer Genome Atlas (TCGA). This data brings with itself significant challenges in the identification of gene expression mechanisms, the determination of proteins encoded by the genes, understanding how these interact; i.e., gene regulatory networks, and marker identification. 

BISP has contributed to this area by introducing a new line of research: genomic and proteomic signal processing \cite{schonfeld2008}. While  biomolecular sequence analysis has been addressed by 
computer scientists, physicists, and mathematicians, it was only at the turn of the century that signal processing has started to play a role in this area. Genomic and proteomic data can be modeled as noisy, continuous or discrete
signals that represent the molecular
structure and activities in cells. The high dimensionality, variability, and complexity of this data require  the development of
new signal processing methodologies that effectively deal
with these challenges.  By mapping the character strings corresponding to gene sequences into numerical sequences, signal processing
offers a set of tools for solving
highly relevant problems. 
For example, the magnitude and the phase of properly defined
Fourier transforms can be used to predict important properties of  protein
coding regions in DNA.  Similarly, concepts from digital filtering can be employed to analyze the mapping of
DNA into proteins and the interdependence of two
sequences. These and other signal processing methodologies, such as frequency domain analysis, high-dimensional data analysis, compressed sensing, and network inference, have played important roles in the advancement of this field. Genomic and proteomic signal processing have had a substantial impact on different applications areas, including sequence analysis, microarray analysis, structure identification and regulatory networks. 

From 2005 to 2013, the IEEE International Workshop on Genomic Signal Processing and Statistics (GENSiPS) was organized annually and sponsored by IEEE SPS. These workshops covered topics related to high-dimensional genomic data analysis, gene regulatory network inference, marker identification, drug screening and proteomics.

\section{Methodological Advances in Biomedical Image and Signal Processing}
The field of biomedical image and signal processing has seen major methodological advances, not only in how data are recorded, stored, and transmitted, but also in how they are best represented, processed, analyzed, and modeled, depending on the application domain. Many paradigms have been proposed in recent decades by various schools of thought, resulting in a wide range of theories and methods for challenging problems such as image and signal restoration, reconstruction, detection, segmentation, classification, pattern recognition, and statistical analysis, as documented in numerous textbooks and reviews. Given the limited space in this article, we only briefly discuss some of the most impactful developments in recent years, including methods for computational imaging, deep learning-based image and signal analysis, and efforts to stimulate reproducible research.

\subsection{Biomedical Computational Imaging}

\begin{figure}[t]
\centering
\begin{tabular}{@{}>{\centering}p{0.25\textwidth}@{}>{\centering}p{0.25\textwidth}@{}>{\centering}p{0.25\textwidth}@{}>{\centering}p{0.25\textwidth}@{}}
    (a) Ground Truth & (b) SNR = 24.06 dB & (c) SNR = 29.06 dB & (d) SNR = 35.38 dB \tabularnewline
    \multicolumn{4}{@{}c@{}}{\includegraphics[width=\textwidth]{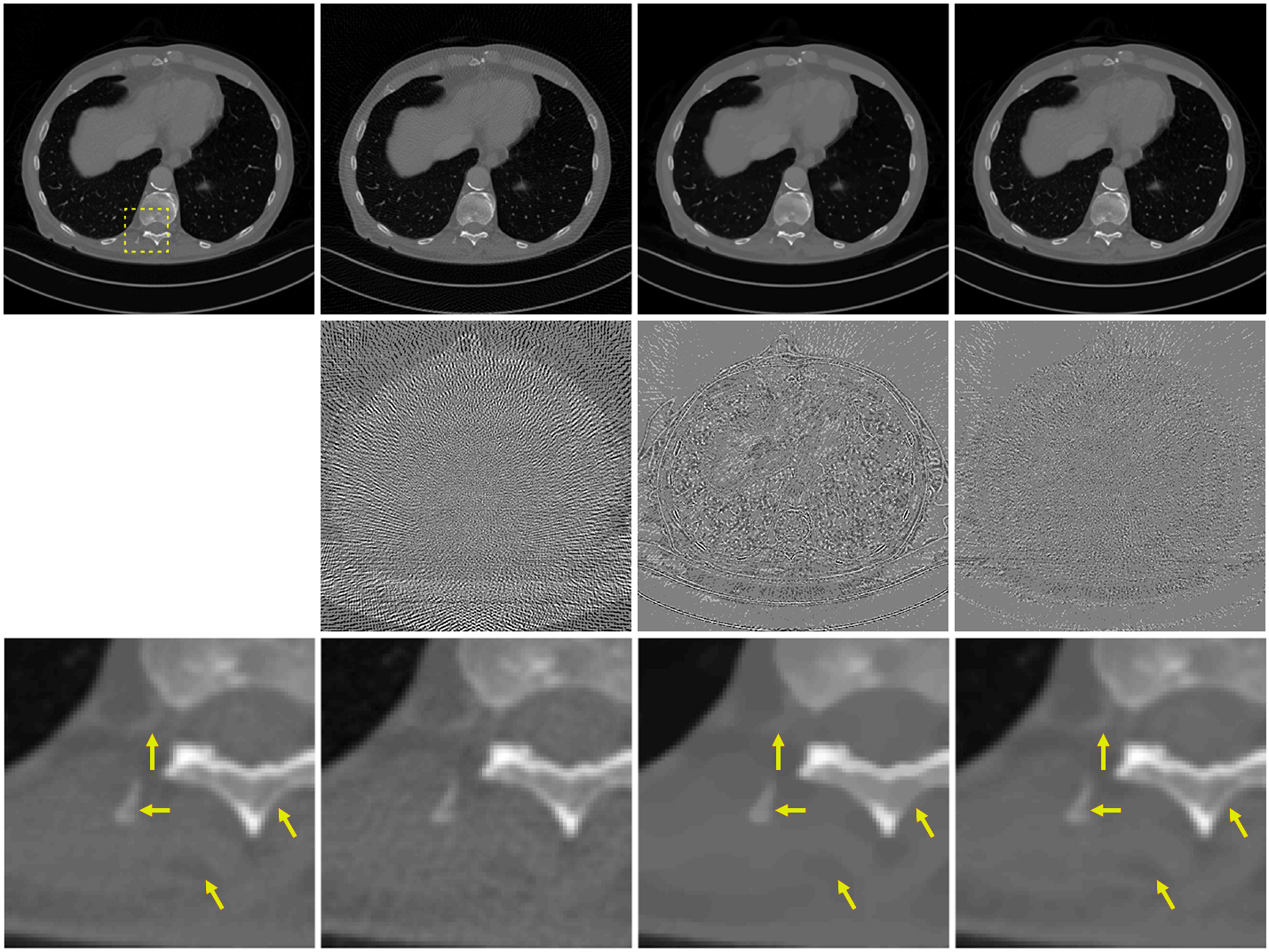}}
\end{tabular}
\caption{Comparison of tomographic reconstruction algorithms for compressed sensing (CS) with a reduction of the number of views by 7. (a) Ground truth (high quality reconstruction from 1000 views). (b) Conventional reconstruction (filtered backprojection) from a subset of 143 views. (c) CS reconstruction using total variation regularization. (d) CS reconstruction using a convolution neural network (FBPConvNet). {\color{red}The middle panel displays the image residuals with the same contrast.} The magnified images in the lower panel represent the corresponding ROI overlaid in (a). The figure is adapted from \cite{Jin2017}.}
\label{fig:Reconstructions}
\end{figure}

Most biomedical imaging modalities have a strong computational component, as they systematically rely on signal processing to reconstruct the images from the raw imaging data. The data can take the form of: (i) 2D projections of a 3D object as in X-ray tomography, PET, and cryo-electron microscopy, (ii) a series of blurred 2D slices of a specimen as in fluorescence microscopy, or (iii) samples of the Fourier transform of an object, as in MRI and optical diffraction tomography. By capitalizing on the knowledge of the imaging physics (linear forward model), the reconstruction task can then be formulated as an inverse problem. Until recently, classical imaging (MRI, CT) relied on a direct inversion of this forward model. This is achieved, for instance, by inverse Fourier transformation in MRI (with uniform sampling in k-space) or by inverse Radon transformation (the celebrated filtered-back-projection algorithm) in CT. This works well when the measurements are sufficiently numerous and diverse, and when the noise is negligible. Besides streamlining of the reconstruction process itself (improved non-uniform (NU) fast Fourier transform (FFT), optimization of sampling parameters, etc.), the earlier involvement of the SP community was to combat the effect of noise with the help of advanced statistical methods. One notable example of such success is the method of ordered subsets in PET and SPECT \cite{Ahn2003}. Another fruitful approach inspired by Wiener filtering is to inject prior information in a stochastic model (e.g., generalized Gaussian in a transformed domain), which makes a direct link between maximum a posteriori (MAP) reconstruction and regularization/energy minimization techniques \cite{Bouman1993}. The more significant revolution in imaging came with compressed sensing (CS) with theorists \cite{Donoho2006,Candes2008} and then experimentalists \cite{Lustig2007,Guerquin2011} showing the feasibility of image reconstruction from a reduced set of measurements. A milestone in this line of research was the development of efficient minimization methods under sparsity constraints, in particular the (fast) iterative soft thresholding algorithm (ISTA) and alternating direction method of multipliers (ADMM) \cite{Figueiredo2003}. The main benefit of CS is to enable faster imaging, which reduces not only cost but also radiation exposure (in the case of X-ray or PET/SPECT). This has led to a major revolution in MRI, with fast (CS-based) imaging protocols now offered by most vendors of MRI technology.  While CS kept SPS researchers busy from 2005-2017, another wave then overtook the field---the incorporation of neural networks in the image reconstruction pipeline. This led to further significant improvement in image quality (Figure~\ref{fig:Reconstructions}), especially in extreme scenarios, e.g., low signal-to-noise ratio and CS \cite{Jin2017}. 
While image reconstruction based on convolutional neural networks (CNNs) still has shortcomings---they are poorly understood and can behave erratically (lack of stability, hallucination)---they demonstrate the potential of better reconstruction quality~\cite{Wang2018}. It is noteworthy that it took SP pioneers less than a year to tune their new CNN-based methods to the point where they would outperform sparsity-based methods for CS in public imaging challenges by the same margin (typically $>4$dB) than the latter had achieved over classical reconstruction during a whole decade of intense research activity. {\color{red} %This explains why 
CNNs and learning-based techniques are presently at the center of attention of the researcher community.
Recent trends include the development of more sophisticated iterative reconstruction schemes that rely on CNNs to regularize the solution---as enabled by the Plug-and-play framework \cite{kamilov2023plug}---as well as the use of deep learning for the resolution of more challenging non-linear inverse problems such as diffuse optical tomography (DOT) \cite{yoo2019deep} and diffraction tomography.}
 
\subsection{Deep Learning in Biomedical Image and Signal Processing}
Traditionally, methods for image and signal processing have been based on carefully designed mathematical models of the phenomena and anomalies of interest, and their translation into efficient rules-based computational algorithms. Illustrative examples of this are mathematical point-spread function (PSF) models of widefield or confocal microscopes based on physical (optical) principles, serving as the basis for various image restoration methods (in particular deconvolution) \cite{Sarder-2006} and object detection methods (such as single molecule localization) \cite{Vonesch-2006}. However, as in many other fields, the demand for new and better methods from practitioners in biology and medical diagnostics outstrips the supply of researchers and developers in image and signal processing. That is, there are many more biologists and physicians in the world looking for tools to facilitate their data processing workflows than there are scientists and engineers looking to develop mathematical models and image/signal processing algorithms specifically for biomedical applications. Moreover, especially in the biomedical field, many image and signal analysis tasks are notoriously difficult to model mathematically, due to the complex nature of the problem, the high ambiguity of the data, and the subjectivity of human experts who define the gold standards for interpreting the data. Thus, as imaging and measurement devices improved over the years, and the number of potentially automated data processing tasks grew, the need for more generic, data-driven, learning-based methods also increased.

\begin{figure}[t]
\centering
\includegraphics[width=\textwidth]{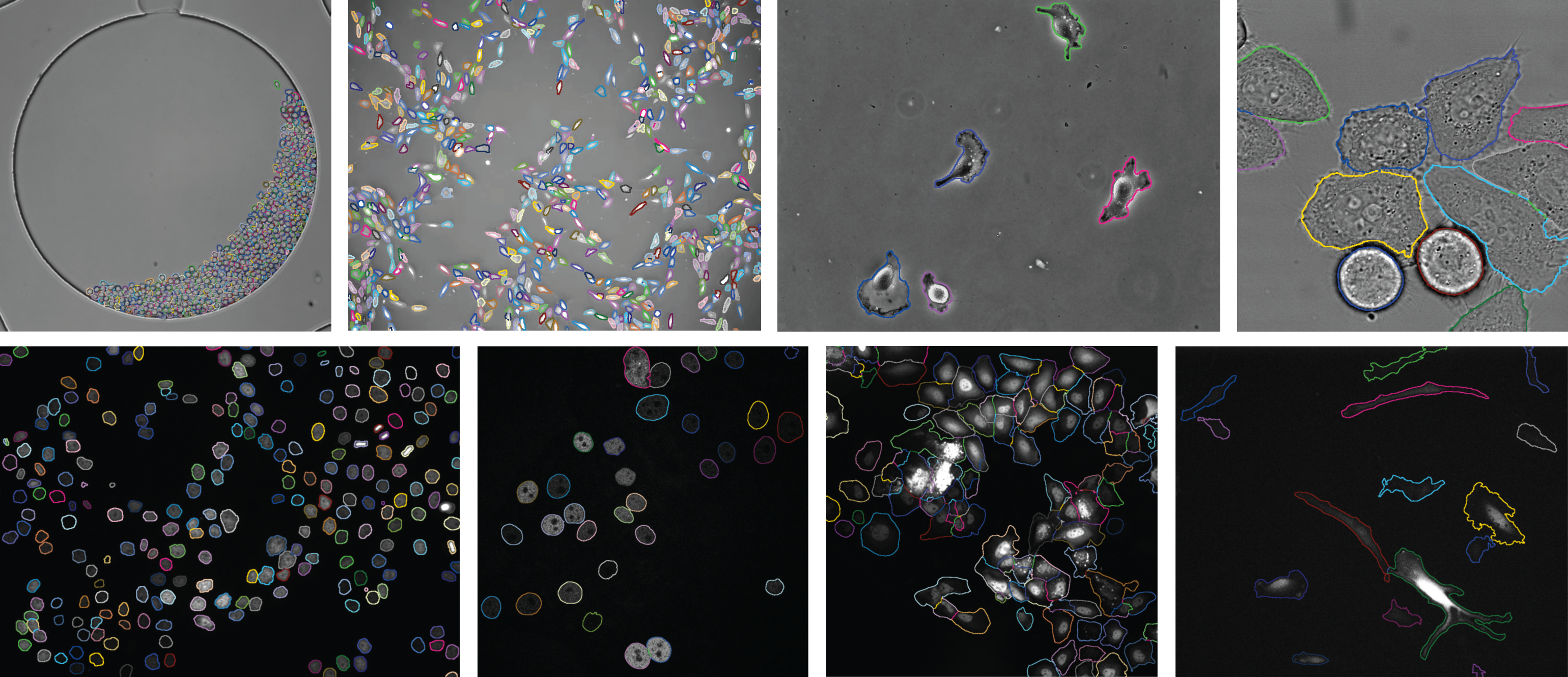}
\caption{Examples of cell segmentation using deep neural networks in diverse types of microscopy images. From top-left to bottom-right, the shown images were captured using bright-field microscopy, phase-contrast microscopy ($2\times$), differential interference contrast (DIC) microscopy, and fluorescence microscopy ($4\times$), and contain distinct types of cells in different spatial arrangements (densities and confluences). The segmentation results are the overlaid colored cell contours (arbitrary colors). These results were produced using a single deep learning framework with a U-Net-like macro-architecture consisting of various layers/blocks whose micro-architectures were optimized automatically using a neural architecture search approach \cite{Zhu-2021}. The examples illustrate the power of deep learning and the level of automation that can be achieved nowadays in optimizing image segmentation results without requiring expert user input, other than manual annotations, to learn from.}
\label{fig:deepsegmentation}
\end{figure}

In the past decade, the rapidly growing availability of large datasets, powerful computing capabilities, and open-access software libraries and frameworks have accelerated the development and adoption of machine learning and deep learning methods in biomedical image and signal processing \cite{Greenspan-2016, Meijering-2020, Zhou-2021, meijering2022deep}. These methods show increasingly superior performance in benchmarking studies for various tasks, including reconstruction, restoration, detection, segmentation, classification, and tracking. In particular, deep learning of artificial neural networks has become a popular approach for solving data analysis problems where multimodal, multidimensional, multiparametric datasets need to be jointly processed, posing a clear challenge to traditional analysis methods. For the processing of biomedical images, especially CNNs have become mainstream, a prominent example of this being the U-Net architecture \cite{Ronneberger-2019}, of which many variants exist for various tasks and applications, such as segmentation (Figure~\ref{fig:deepsegmentation}). For biomedical signal processing, especially for dealing with time series, recurrent neural networks (RNNs) such as the long short-term memory (LSTM) unit have seen widespread adoption. However, despite promising results, many challenges remain to be addressed before deep learning solutions can be integrated with full confidence and accountability into the workflows of biomedical practitioners\red{, such as developing ways to incorporate expert knowledge and improving the explainability and generalizability of the models (see the discussion of future directions in the last section)}.

\subsection{Reproducible Research, Open Access, Code}
Reproducing the results presented in a research paper can be very challenging. For a computational algorithm, details such as the exact dataset, initialization or termination procedures, and precise parameter values are often omitted in the publication for various reasons. This makes it difficult, if not impossible, for someone else to obtain the same results \cite{vandewalle2009reproducible}. In the early 2000s, the need to boost research by implementing reproducible research practices became apparent. Vandewalle et al. \cite{vandewalle2009reproducible} published a seminal manuscript in the IEEE Signal Processing Magazine in 2009, which defines reproducible research: ``A research work is called reproducible if all information relevant to the work, including, but not limited to text, data, and code, is made available, so that an independent researcher can reproduce the results.'' The authors also distinguish six levels of reproducibility, from Level 5 (an independent researcher can easily reproduce results with at most 15 minutes of user effort, requiring only standard, freely available tools---C compiler, etc.) to Level 0 (an independent researcher can reproduce results).  

\begin{figure}[t]
\centering
\includegraphics[angle=90,width=\textwidth]{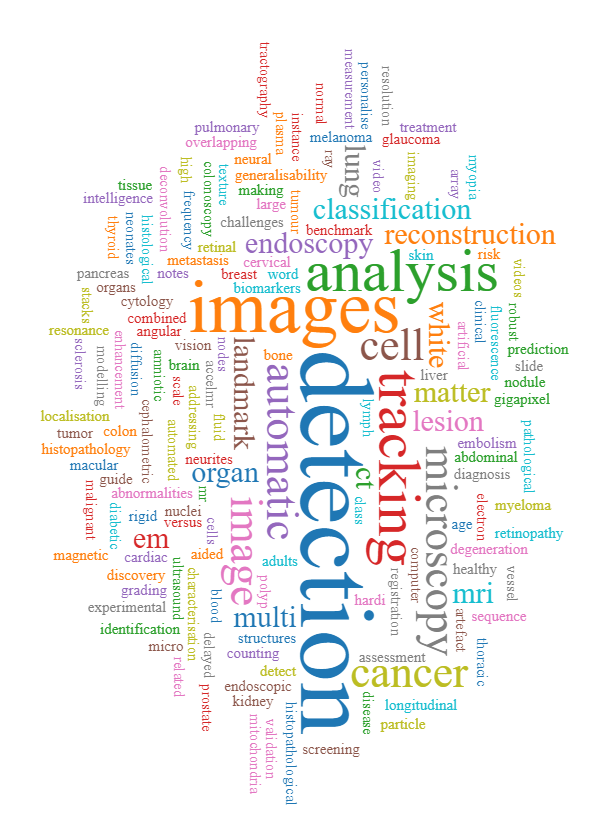}
\caption{The IEEE International Symposium on Biomedical Imaging (ISBI) is the premier scientific venue for the BISP TC. Since 2012, our community has organized over 60 challenges, where open datasets, well-specified tasks, and evaluation metrics have been made available for multiple groups to participate, compete, and learn from each other. Challenges have covered many imaging modalities and scales, image computing tasks, and organ systems.}
\label{fig:isbi-challenges}
\end{figure}

The issue of reproducibility has been raised many times in the past few decades. In the 1980s, there was a growing awareness of poor building on the previous work of others \cite{Price1986}. Published algorithms were frequently evaluated, with only select data and \emph{ad hoc} metrics. The comparison of algorithms and software performance was difficult. In the 1990s, the first benchmark initiative in biomedical imaging, ‘Retrospective Image Registration Experiment (RIRE)’, appeared at Vanderbilt University \cite{west1997comparison}. We needed to wait until the early 2010s for bioimaging reference datasets and challenges (benchmarks associated with competitions) to appear. ISBI 2012 in Barcelona was the first edition of the symposium holding challenges, on the following topics: 1) Particle tracking \cite{ptc2014}; 2) Segmentation of neuronal structures in EM stacks; 3) Vessel segmentation in the lung \cite{rudyanto2014comparing}; 4) Cardiac delayed-enhancement magnetic resonance image segmentation; 5) High angular resolution diffusion imaging; and 6) Challenge US: Biometric measurements from fetal ultrasound images. At ISBI 2015, Prof. Ronneberger's team won the Cell Tracking Challenge (third edition) \cite{mavska2014benchmark} and the dental X-ray image segmentation challenge with their U-Net \cite{Ronneberger-2019}. Figure \ref{fig:isbi-challenges} shows a word cloud of the challenge titles over the years; detection, images and tracking occupy a prominent place in the cloud.

In the bioimaging community, the push for reproducibility led to several open software platforms, such as Cell Profiler (https://cellprofiler.org/), Fiji (https://fiji.sc/) and Icy (https://icy.bioimageanalysis.org/). They were made available to the community in the early 2000s to share the then state-of-the-art analytical methods, which are now used for integrated deep learning framework deployment \cite{gomezdemariscal}.

Finally, imaging challenges foster collaboration between institutions and continents. Since 2012, over 60 challenges have been organized, led by multiple institutions. Of these, 31 were organized by European institutions, 12 by organizations in the Americas, 10 by Asia or Oceania, and 6 involving cross-continental collaboration from the Americas, Asia, and Europe. These collaborations drove our community to learn from the strengths and pitfalls\cite{Reinke2021} in organizing challenges and interpreting their results \cite{Maier_Hein-2018} and thus developed best-practice guidelines for transparent reporting \cite{Maier_Hein-2020}. 

\section{Future Directions}
Advanced technologies for capturing biomedical images and signals have made a growing and lasting positive impact on clinical diagnostics and therapeutics, medical research, and life sciences. They will continue to help improve our understanding of the conditions underlying human health and how to prevent and treat disease. Modern biomedical image and signal acquisition systems are based on a wide range of physical phenomena (electricity, magnetism, light, sound, force, etc.) capable of providing complementary information about the anatomical and functional properties of the human body and living organisms in general. Also, the sensitivity, resolution, and quality of these systems have improved dramatically over the years, to the point where automated image and signal processing are now indispensable in virtually all clinical and biomedical research applications. At this point in time, unlike in the past century, the chances of discovering totally new physical principles that could ultimately be used in biomedical practice have diminished, yet the challenges of fully exploiting existing technologies are far from having been solved. One of the main problems for the image and signal processing community in the years ahead will be to develop effective methods for data fusion and integration \cite{Lahat-2015} to maximize the potential of multimodal and correlative imaging, as well as combining imaging and non-imaging (e.g. ``omics’’) data. This requires finding solutions to dealing with the fundamentally different nature of different data sources, and the inevitable imbalances in the data, but also with the huge volumes (terabytes and no doubt soon petabytes) of multimodal datasets.

Despite being comparatively young, the BISP community has already seen and contributed to major paradigm shifts in biomedical image and signal processing. Still, in addition to the data challenges mentioned above, many fundamental technical challenges remain. Examples include some of the problems caused by the increasing emphasis on learning-based approaches. For starters, these approaches are typically very data-hungry, while the human and time resources to produce high-quality annotated datasets are usually severely limited, especially in the biomedical domain, not to mention additional limiting factors due to privacy regulations. This requires the development of semi/unsupervised learning approaches, data modeling and simulation methods that can generate high-fidelity ground-truth data for training, and ways to integrate expert domain knowledge into the learning framework. Furthermore, even if sufficient annotated data can be collected to train a machine or deep learning-based method for a given application, the resulting model is considered a black box in the eyes of practitioners, who remain fully accountable for any decisions based on the model’s predictions. Hence, there is a great need for explainable and interpretable machine and deep learning solutions. This is a fantastic opportunity for BISP researchers, many of whom traditionally are used to developing mathematical models based on sound physical principles, which by design are much more explainable and interpretable. Another challenge stemming from limited training data is the typically poor generalizability of the learned models. While organized competitions in the field have done a great service by providing public datasets and benchmarks, it is now well-known that models based on them do not always work on private datasets. This calls for continuing efforts to make public datasets less selective and more representative.

Given these and many other open challenges, the BISP TC will continue to play an important role in developing ever more advanced image and signal processing methodologies underpinning the next generation technologies needed to improve the efficacy of biomedical practice and research. In this endeavor, we believe future advances will not only come from continuing research efforts, but also from innovations in education and how we train the next generation of scientists and engineers in our field. Clearly, biomedical image and signal processing has become increasingly multidisciplinary, requiring a deep understanding of not only the mathematics and algorithms of how to model and process digital images and signals, but also of the underlying physical principles and limitations of data acquisition using various systems, the biomedical knowledge to properly interpret the data, the data science and informatics expertise to handle large datasets, and the experimental and statistical know-how to validate methods thoroughly. To this end, we envision the BISP TC to strengthen ties with the relevant bodies in the respective disciplines, and to become more multidisciplinary in the future.

%caution about AI in bioimaging, explainability
%education (a paragraph), need for math, understanding of physics, careful design of experiments and validation;
%understanding the underlying physiology and biology

%\section{References to use}
%\cite{adali2008introduction,dougherty2005genomic,farina2021signal,kervrann2016,kovacevic2006,meijering2022deep,munoz2015quantitative,murphy2005,unser2002guest,van2016introduction,vandewalle2009reproducible,wang2018image,adali2013,monga2020,schonfeld2008,unser2003guest}

\bibliographystyle{IEEEtran}
\bibliography{refs}

\end{document}